\documentclass[aps,superscriptaddress,showpacs,floatfix,amsmath,amssymb,twocolumn]{revtex4}

\usepackage{epsfig}
\usepackage{dcolumn}
\usepackage{bm}
\newcommand{\nuc}[2]{\hbox{$^{#1}$#2}}
\begin{document}

\title{Spectroscopy of $^{35}$P using the one-proton knockout reaction}

\author{A.~Mutschler}
\affiliation{Institut de Physique Nucl\'eaire, IN2P3-CNRS,
F-91406 Orsay Cedex, France}
\affiliation{Grand Acc\'el\'erateur National d'Ions Lourds (GANIL),
CEA/DSM - CNRS/IN2P3, B.\ P.\ 55027, F-14076 Caen Cedex 5, France}

\author{O.~Sorlin}
\affiliation{Grand Acc\'el\'erateur National d'Ions Lourds (GANIL),
CEA/DSM - CNRS/IN2P3, B.\ P.\ 55027, F-14076 Caen Cedex 5, France}

\author{A.~ Lemasson}
\affiliation{Grand Acc\'el\'erateur National d'Ions Lourds (GANIL),
CEA/DSM - CNRS/IN2P3, B.\ P.\ 55027, F-14076 Caen Cedex 5, France}
\affiliation{Department of Physics and Astronomy and National
Superconducting Cyclotron Laboratory, Michigan State University,
East Lansing, Michigan, 48824-1321, USA}

\author{D. Bazin}
\affiliation{Department of Physics and Astronomy and National
Superconducting Cyclotron Laboratory, Michigan State University,
East Lansing, Michigan, 48824-1321, USA}

\author{C.\ Borcea}
\affiliation{IFIN-HH, P. O. Box MG-6, 76900 Bucharest-Magurele, Romania}

\author{R.\ Borcea}
\affiliation{IFIN-HH, P. O. Box MG-6, 76900 Bucharest-Magurele, Romania}


\author{A. Gade}
\affiliation{Department of Physics and Astronomy and National
Superconducting Cyclotron Laboratory, Michigan State University,
East Lansing, Michigan, 48824-1321, USA}


\author{H. Iwasaki}
\affiliation{Department of Physics and Astronomy and National
Superconducting Cyclotron Laboratory, Michigan State University,
East Lansing, Michigan, 48824-1321, USA}

\author{E. Khan}
\affiliation{Institut de Physique Nucl\'eaire, IN2P3-CNRS,
F-91406 Orsay Cedex, France}

\author{A.~Lepailleur}
\affiliation{Grand Acc\'el\'erateur National d'Ions Lourds (GANIL),
CEA/DSM - CNRS/IN2P3, B.\ P.\ 55027, F-14076 Caen Cedex 5, France}

\author{F. Recchia}
\affiliation{Department of Physics and Astronomy and National
Superconducting Cyclotron Laboratory, Michigan State University,
East Lansing, Michigan, 48824-1321, USA}

\author{T. Roger}
\affiliation{Grand Acc\'el\'erateur National d'Ions Lourds (GANIL),
CEA/DSM - CNRS/IN2P3, B.\ P.\ 55027, F-14076 Caen Cedex 5, France}

\author{F.~ Rotaru}
\affiliation {IFIN-HH, P. O. Box MG-6, 76900
Bucharest-Magurele, Romania}

\author{M.\ Stanoiu}
\affiliation {IFIN-HH, P. O. Box MG-6, 76900
Bucharest-Magurele, Romania}

\author{S. R. Stroberg}
\affiliation{Department of Physics and Astronomy and National
Superconducting Cyclotron Laboratory, Michigan State University,
East Lansing, Michigan, 48824-1321, USA}
\affiliation{TRIUMF, 4004 Westbrook Mall, Vancouver, British Columbia, V67 2A3 Canada}

\author{J.\ A.\ Tostevin}
     \affiliation{Department of Physics, University of Surrey,
       Guildford, Surrey GU2 7XH, United Kingdom}

\author{M. Vandebrouck}
\affiliation{Institut de Physique Nucl\'eaire, IN2P3-CNRS,
F-91406 Orsay Cedex, France}
\affiliation{Grand Acc\'el\'erateur National d'Ions Lourds (GANIL),
CEA/DSM - CNRS/IN2P3, B.\ P.\ 55027, F-14076 Caen Cedex 5, France}

\author{D. Weisshaar}
\affiliation{Department of Physics and Astronomy and National
Superconducting Cyclotron Laboratory, Michigan State University,
East Lansing, Michigan, 48824-1321, USA}

\author{K.~Wimmer}
\affiliation{Department of Physics, The University of Tokyo, Hongo, Bunkyo-ku, Tokyo 113-0033, Japan}
\affiliation{Department of Physics, Central Michigan University, Mt. Pleasant, Michigan 48859, USA}
\affiliation{Department of Physics and Astronomy and National
Superconducting Cyclotron Laboratory, Michigan State University,
East Lansing, Michigan, 48824-1321, USA}

\begin{abstract}

The structure of $^{35}$P was studied with a one-proton knockout reaction at
88~MeV/u from a $^{36}$S projectile beam at NSCL. The $\gamma$ rays from the
depopulation of excited states in $^{35}$P were detected with GRETINA, while
the $^{35}$P nuclei were identified event-by-event in the focal plane of the
S800 spectrograph. The  level scheme of $^{35}$P was deduced up to 7.5 MeV using
$\gamma-\gamma$ coincidences. The observed levels were attributed to proton
removals from the $sd$-shell and also from the deeply-bound $p_{1/2}$ orbital.
The orbital angular momentum of each state was derived from the comparison
between experimental and calculated shapes of individual ($\gamma$-gated)
parallel momentum distributions. Despite the use of different reactions and
their associate models, spectroscopic factors, $C^2S$, derived from the
$^{36}$S $(-1p)$ knockout reaction agree with those obtained earlier from
$^{36}$S($d$,\nuc{3}{He}) transfer, if a reduction factor $R_s$, as deduced
from inclusive one-nucleon removal cross sections, is applied to the knockout transitions.
In addition to the expected proton-hole configurations, other states were observed
with individual cross sections of the order of 0.5~mb. Based on their shifted
parallel momentum distributions, their decay modes to negative parity states,
their high excitation energy (around 4.7~MeV) and the fact that they were not
observed in the ($d$,\nuc{3}{He}) reaction, we propose that they may result
from a two-step mechanism or a nucleon-exchange reaction with subsequent neutron
evaporation. Regardless of the mechanism, that could not yet be clarified, these
states likely correspond to neutron core excitations in \nuc{35}{P}. This
newly-identified pathway, although weak, offers the possibility to selectively
populate certain intruder configurations that are otherwise hard to produce
and identify.

\end{abstract}

\date{\today}
\maketitle

\section{Introduction.}

For many decades, single nucleon transfer reactions have been one of the tools
of choice for the study of shell structure in nuclei. Various combinations of
light projectiles were used on stable targets to probe occupied single-particle
levels and valence states, e.g. the $(d,p)$ neutron-adding transfer to study the
valence neutron shell and the ($d,^{3}$He) proton-removing transfer to study
occupied proton orbitals. At facilities where radioactive ion beams are available
with sufficient intensity, this approach is also now used for unstable nuclei
in inverse kinematics, probing those active orbitals near the Fermi surface.
Separation energies and orbital angular momenta and the occupation of nucleon
orbitals in the nucleus of interest are extracted from the energy and angular
distributions of the light ejectile and from the sum of partial cross sections,
respectively. Using direct reaction theory, this is done in a model-dependent way,
including the use of appropriate optical potentials in the entrance and exit
channels. The magnitudes and evolution of shell gaps and the softness or stiffness
of the nuclear Fermi surfaces were studied on many stable (and recently also on
a few unstable) nuclei using these techniques (see e.g. Refs~\cite{UozPRC,UozNPA}
for results on the stable $^{40}$Ca and $^{48}$Ca nuclei).

Even for closed-shell nuclei, short-range correlations~\cite{Pand97} and coupling
to collective degrees of freedom \cite{Barb09} complicate the determination of
single-particle energies and spectroscopic factors (or their related shell
occupancies and vacancies) which are not directly observable \cite{Dick04,Dugu15}.
Furthermore, reaction models exploit {\em effective} potentials that do not capture
the full microscopic complexity of the nucleus and shell-model interpretations can
depend strongly on truncations of the valence space. As a result, values of
spectroscopic factors, $C^2S$, deduced experimentally are quenched compared to
theoretical calculations. This quenching is typically a factor of about
$R_s=0.55(10)$ \cite{Lapi93,Kay13} in stable and near-stable nuclei.

Strictly speaking, the only {\it true} experimental observables are the energies
of levels in the final nucleus and their partial feeding cross sections. However,
to interpret the underlying nuclear structure it is important
to infer shell evolution and occupancies from these observables. Assuming collective
coupling (e.g. to giant resonances) and short-range correlation effects are similar
between neighboring nuclei, consideration of the {\it differential} evolution of
single-particle energies, spectroscopic strengths and occupancies is sensible,
a view supported by their successful use up to now.

With the development of accelerator facilities producing intense high-energy
secondary beams of rare isotopes at energies of around 100 MeV/u or more, the
linear and angular momentum matching of light-ion-induced pick-up
or stripping reactions is poor, resulting in small cross sections of the order
of 1~mb. With thin targets, this technique is then only applicable for the most
intense beams. A higher-luminosity alternative, the intermediate-energy nucleon
knockout reaction, is also being exploited successfully to study the spectroscopy
of extremely neutron-rich or deficient nuclei. Being sensitive to the same single-
nucleon overlaps, empirical proton and neutron spectroscopic factors $C^2S(\alpha)$
to bound states $\alpha$ in the residual, mass $(A-1)$, nucleus can be deduced from
partial cross-section measurements~\cite{Hansen2003,Tostevin99}. For 
near-stable nuclei with modest proton to neutron separation energy asymmetry, 
i.e. $|\Delta S|= |S_p - S_n|< 10$ MeV, fast proton knockout, ($d,^{3}$He) and 
electron-induced proton knockout $(e,e',p)$ reactions have all been used
to populate proton-hole-like final states by proton removal. In each case, reaction 
model calculations that use shell-model $C^2S$ must be suppressed by very similar 
factors, $R_s$, to agree with data \cite{Kram01,Kay13,Tost14}; this despite the very 
different reaction energies, mechanisms, models and approximations involved.
In the nuclear knockout case, this suppression has been quantified for many systems
at the level of the inclusive cross sections, the summed partial cross sections
to all bound final states of the reaction residues. Based on a compilation of this
large body of experimental one-proton and one-neutron knockout reactions data on
both neutron-rich and neutron-deficient nuclei, a significant variation of the
$R_s=\sigma_{exp}/\sigma_{th}$ value (from 0.3 to 1) was observed when systems
with extreme $\Delta S$ values were explored \cite{Tost14}. So far, this strong
trend has not been reported when using $(p,d)$ or ($d,^{3}$He) reactions in the
few systems studied with significant $\Delta S$ asymmetry \cite{Lee10,Flav13},
potentially the result of large uncertainties in the transfer reaction theory
\cite{Nunes2011} for such highly-mismatched transitions.

An aim of the present work is to compare the states populated, their deduced
angular momenta and their $C^2S$ values as extracted from the one-proton knockout
reaction $^{36}$S$(-1p)$$^{35}$P with those from the $^{36}$S($d,^{3}$He)$^{35}$P
stripping reaction \cite{Khan85}. This study covers proton separation energies
that differ by 10~MeV, from the removal of the most deeply-bound proton from
the $1p_{1/2}$ orbital to removal of a $1d_{3/2}$ valence proton from $^{36}$S.
This work shows that the one-proton knockout and ($d,^{3}$He) transfer reactions
can be analyzed to extract consistent spectroscopic information. Furthermore, a
few percent of the observed cross section leads to suspected negative parity 
states that cannot be interpreted as due to the direct one-proton knockout 
reaction. Their origin will be discussed in the final part of the manuscript.

\section{Experimental procedure}
A secondary beam of $^{36}$S was produced in the fragmentation of a 140 MeV/u
$^{48}$Ca primary beam on a 846 mg/cm$^2$ $^9$Be target, delivered by the
coupled cyclotron facility at NSCL. The  $^{36}$S nuclei were
selected with the A1900 fragment separator~\cite{Stolz05,Morrissey03}, yielding
an average intensity and purity of 8.1 $10^5$ s$^{-1}$ and 89.7\%,
respectively. The constituents in the  projectile beam were identified from
their time-of-flight difference provided by two plastic scintillators located
before the secondary target. The 100-mg/cm$^{2}$ secondary target of
$^9$Be was located at the reaction target position of the S800 spectrograph
whose magnetic rigidity was centered on the $^{35}$P residues produced in the
one-proton knockout from the \nuc{36}{S} projectiles. The projectile-like
reaction residues
were identified on an event-by-event basis from their energy loss measured in an
ionization chamber located at the focal plane of the S800 and from their
time-of-flight measured between two scintillators situated at the object
position  and at the focal plane of the  S800 spectrometer. The trajectories of
the residues were determined from positions and angles determined in the focal
plane using two cathode-readout drift
chambers \cite{Yurkon99}. Their  non-dispersive position and the momentum vector
at the target position  were reconstructed using the ray-tracing ion-optics
code COSY~\cite{Berz93}. A total of  3.7 $\cdot$ 10$^6$ $^{35}$P nuclei produced
from \nuc{36}{S} were registered during the measurement.

Prompt $\gamma$ rays corresponding to the deexcitation of the $^{35}$P
residues in flight were detected with the seven modules of the Gamma-Ray Energy
Tracking In-beam Nuclear Array (GRETINA)~\cite{Paschalis13} that surrounded the
secondary target position. Four of the detectors covered the most forward angles
centered on 58$^\circ$, while the remaining three were installed at 90$^\circ$.
Event-by-event Doppler correction was performed using the event-by-event ion
velocity and position at mid-target, as well as the $\gamma$-ray detection angle
derived from the $\gamma$-ray interaction position in the segments of the
GRETINA array. For this it was assumed that the first interaction point
corresponds to the interaction point with highest energy deposition (main
interaction scheme). The
energies of $\gamma$ rays
detected in the same or neighboring crystals were considered to come from one
single event. They were summed in an add-back procedure to increase the
$\gamma$-ray efficiency in particular at high energy. These add-back spectra
were used also to establish the level scheme  of $^{35}$P using $\gamma$-$\gamma$
coincidence from the full GRETINA array using all angles. Absolute $\gamma$-ray 
intensities and deduced final-state populations were derived without add-back 
from the modules at 90$^{\circ}$ only, since -- at the very high
projectile rates of close to 1~MHz -- the forward detectors were exposed to
frequent high-energy events likely induced by light
particles~\cite{Stroberg14}, causing preamplifier saturation and preventing a
reliable deadtime determination for these detectors.

Gamma-ray efficiencies were determined up to 1.4
MeV using a calibrated $^{152}$Eu  $\gamma$-ray source. Absolute $\gamma$-ray
efficiencies extrapolated to higher energies and that take the Lorentz boost
from the in-flight emission of the $\gamma$-ray into account were obtained from
GEANT4 simulations \cite{Agostinelli03}, yielding simulated efficiencies of
5.5\% (7\%) at 1 MeV without (with) add-back. Gamma-ray energy centroids were
determined with an uncertainty of 2~keV.

\section{Level Scheme of $^{35}$P}

\begin{figure} [h]
\includegraphics[width=\columnwidth] {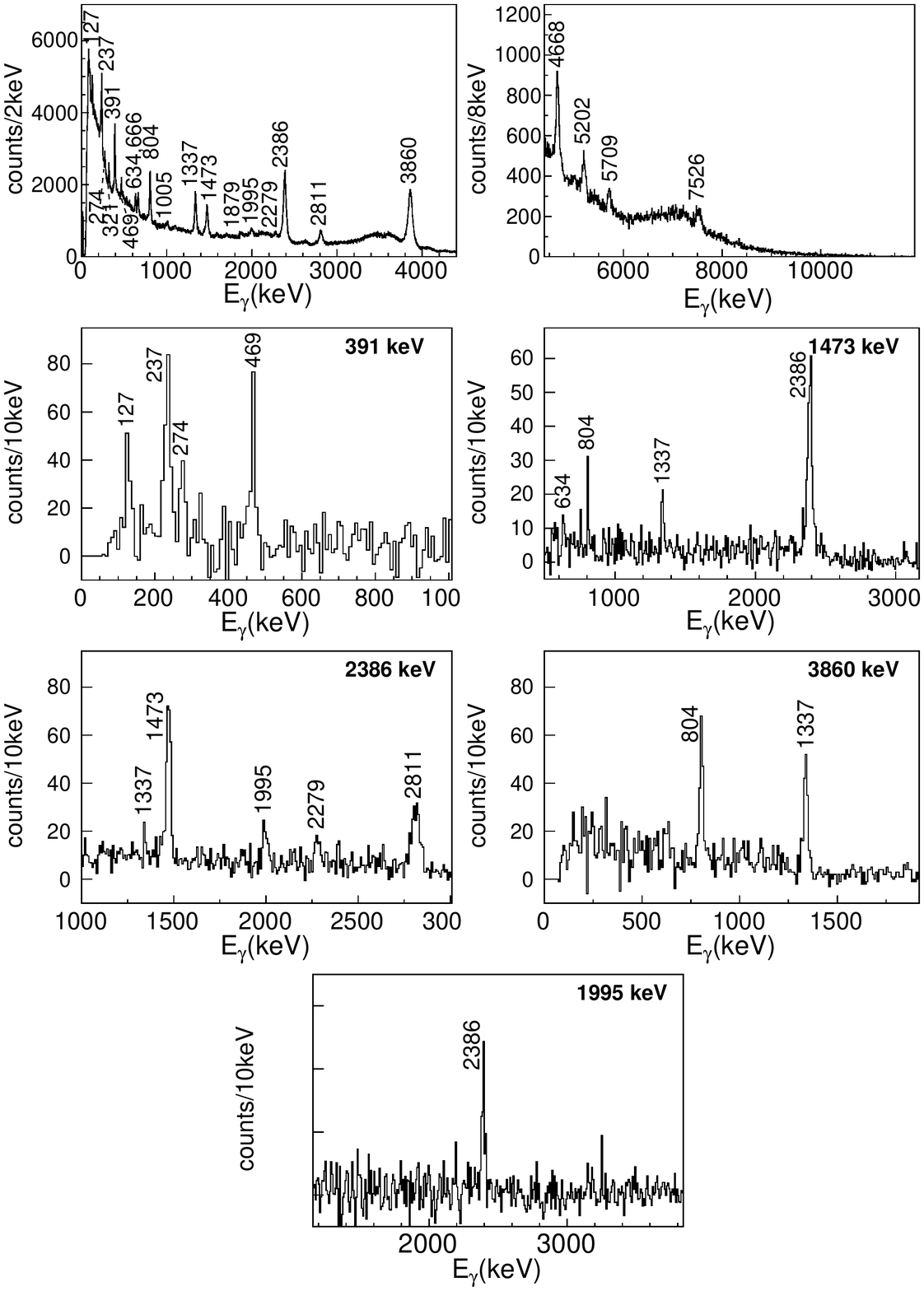}
\caption {Singles (first line) and $\gamma$-gated (second to fourth lines)
$\gamma$-ray Doppler-corrected spectra of $^{35}$P. }
\label{35P-gamma}
\end{figure}

\begin{figure} [ht]
\includegraphics[width=8.5cm] {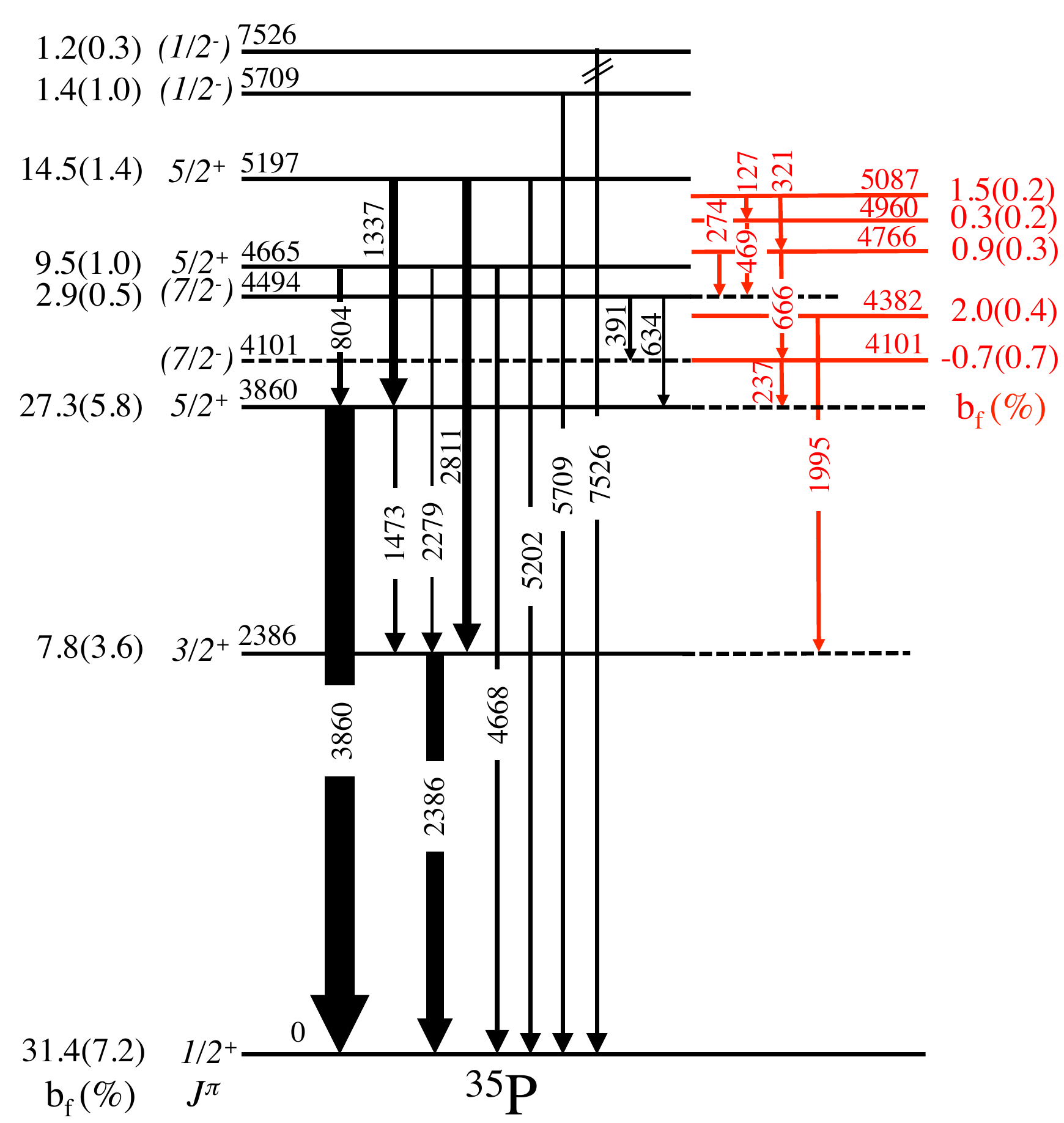}
\caption {(color on line) (a) Energies (with uncertainties of $\sigma$=2 keV) and population fractions, $b_f$ (in $\%$) of the states fed in $^{35}$P during our experiment, by the one proton-knockout reaction (in black), or possibly by a two-step reaction mechanism (right part, in red), see text for details. The reported spins and parities of states, taken from Ref. \cite{Wide08}, are confirmed by the present work.}
\label{35P-ls}
\end{figure}

The Doppler-corrected  singles $\gamma$-ray spectrum of $^{35}$P is shown in the
first row of Fig.~\ref{35P-gamma} for two energy ranges spanning up to 10 MeV. 
Owing to the high $\gamma$-ray detection efficiency, photo peaks are still 
observed up to 7.5 MeV, well above the known $\gamma$ rays observed so far in this
nucleus, and close to the neutron separation energy of $S_n=8.38$ MeV. The
$\gamma$-ray spectra of $^{35}$P, gated on the 391, 1473, 2386, 3860 and 1995 keV
lines, are  shown in the second to fourth rows of Fig.~\ref{35P-gamma}. The
$\gamma$-$\gamma$ coincidences deduced from these spectra have been used to establish
the level scheme of $^{35}$P. The deduced fractional population of each level, $b_f$, 
is given in Fig. \ref{35P-ls} in units of~\%.

Interestingly, all the levels displayed with black lines in Fig.~\ref{35P-ls}
that have a significant feeding were also populated in the
$^{36}$S($d,^{3}$He)$^{35}$P reaction performed in normal kinematics and at low
energy \cite{Khan85}. There, the positive-parity states, the $1/2^+$ (g.s.) and
the $5/2^+$ states (3860, 4665, 5197 keV) were interpreted as hole
configurations originating from the removal of a proton from the occupied
$2s_{1/2}$ and $1d_{5/2}$ orbitals in $^{36}$S, respectively, while the observed
$3/2^+$ (2386 keV) was proposed to correspond to a proton excitation into the
$1d_{3/2}$ orbital. No spin assignment was proposed for the highest energy
levels at 5709~keV (not observed in Ref. \cite{Khan85}) and at 7526 keV, which
we propose to correspond to hole configurations involving the deeply-bound $p_{1/2}$
orbital, as discussed later. Regardless of the spin assignments and spectroscopic
factor values that will be discussed in the next section, this shows that the
($d,^{3}$He) transfer reaction and the one-proton knockout reaction populate the
same states in $^{35}$P. Since derived from high-resolution spectroscopy of
$\gamma$ rays, the energies of these states are more accurate than
the ones from the transfer reaction for which uncertainties of about 20 keV were
reported.

The states at 4101 and 4494 keV were reported in Ref.~\cite{Dufour}
from the $\beta$-decay of $^{35}$Si to $^{35}$P as well as in Ref.~\cite{Wide08}
using the $^{208}$Pb($^{36}$S,X+$\gamma$) reaction at 230 MeV. States at
5560 and 6096 keV that were proposed to be populated in the $\beta$ decay are
neither observed in the present experiment, nor in Ref.~\cite{Wide08}.  As for the
5560 keV state, it is most likely not populated in our experiment since we do
not observe any evidence for the 3174-keV $\gamma$-ray branch deexciting it. The
6096-keV state was proposed to decay with a 1995 keV transition, followed by
cascades containing 241, 1473, 1715, 2386, 3860 and 4101
keV $\gamma$ rays.  In our work, as in Ref. \cite{Wide08}, the 1995-keV
transition is in coincidence with the 2386-keV line but not with any other of
the observed transitions at 237, 1473 and 3860 keV. We therefore suggest that
our reported 4382-keV state (4381 keV in~\cite{Wide08}) is the origin of the
1995-keV $\gamma$ ray transition and that the suggested 6096-keV level reported
from the $\beta$-decay experiment~\cite{Dufour} was likely wrongly placed. In order to
match allowed $\beta$-decay Gamow-Teller selection rules, the states
at 4101, 4382 and 4494 keV, fed from the J$^\pi$=7/2$^-$ g.s. of $^{35}$Si,
should have  5/2$^-$ $\leqslant$ J $^\pi$ $\leqslant$ 9/2$^-$ spin and parity
assignments.

The states at 4382, 4766, 4960 and 5087~keV, indicated in red in
Fig. \ref{35P-ls}, were already identified in Ref.~\cite{Wide08}. They are
populated in our experiment with rather weak individual fractions, $b_f$,
of around 1-2\%, at a total of 4.7\%, but were not observed with the
($d,^{3}$He) transfer reaction \cite{Khan85}. This could either be due to the
better sensitivity of the present experiment as compared to the work of
\cite{Khan85}, or it could point to the fact that these states are not populated
in the one-proton knockout reaction but rather by another process. To shed light
on this unexpected feature, we first note that these states mostly decay by
low-energy $\gamma$ rays to each other or to previously-suggested
negative-parity 7/2$^-$ states at 4101 and 4494~keV rather than by high-energy
transitions to lower-lying positive-parity low-spin states.
One may therefore suspect that they have negative parity or/and spin values
larger or equal to $J=5/2$. As these states are located at energies similar to
the $J^\pi$=3$^-$,4$^-$,5$^-$ states in $^{36}$S that correspond to neutron core
excitations ($d_{3/2}$)$^{-1}$($f_{7/2}$)$^{+1}$, one may suspect that they have
the same origin. These neutron core excitations could not be populated through a
direct one-proton knockout reaction but must rather come from another process, as
will be discussed in Section~\ref{TS}.

\section{Knockout reaction calculations}

The eikonal model and the choice of parameters used to calculate the proton-removal
single-particle cross sections, $\sigma_\alpha^{sp}$, to residue final states $\alpha$
-- the cross sections for removal of a proton with quantum numbers $\alpha$=$n,
\ell,J$ and unit spectroscopic factor -- are detailed in Ref. \cite{Gade08}. These 
$\sigma_\alpha^{sp}$, each the sum of stripping and diffractive removal contributions,
are computed based on the radial overlaps of the initial and final state wave functions 
and the $^{35}$P residue- and proton-target elastic S-matrices calculated from their 
complex optical interactions with the $^9$Be target \cite{Tostevin01}. These S-matrices 
are calculated in the double- and single-folding model (optical) limit of Glauber's 
multiple-scattering theory. The residue-target interaction uses the proton and neutron 
densities of $^{35}$P from Skyrme (SkX interaction~\cite{Brown98}) Hartree-Fock (HF) 
calculations. The $^9$Be target density is a Gaussian with a root-mean-squared (rms) 
radius of 2.36~fm. The removed proton-$^{35}$P relative motion wave functions (radial 
overlaps) and their single-particle rms radii are also constrained, consistently, by 
HF calculations (see \cite{Gade08} for further details).

Systematic analyses have been made of an increasing body of precision inclusive nucleon 
knockout (KO) data by consistent use of this methodology combined with the shell-model 
spectroscopy (levels and spectroscopic factors) appropriate to each case. These
systematics, summarized recently in Ref.~\cite{Tost14}, show a clear correlation of the 
ratio of the experimental to the theoretical (eikonal plus shell-model) inclusive knockout 
cross sections, $R_s=\sigma_{inc,KO}^{exp}/\sigma_{inc,KO}^{th}$, with the asymmetry 
$\Delta S = S_p - S_n$ of the proton ($S_p$) and neutron ($S_n$) separation energies
from the projectile. For proton knockout reactions this dependence can be parameterised 
as \begin{equation}
R_s= -0.016 [ S_p  - S_n ] + 0.61 \label{eq1}
\end{equation}
with an associated error of order 20\%. 

As discussed in the previous section, the measured partial cross section to each 
final state $f$ of $^{35}$P is $\sigma_f=b_f \,\sigma_{inc}^{exp}$, where 95.3\% of 
the total inclusive cross section, $\sigma_{inc}^{exp}$, is attributed to states 
populated by the direct proton knockout mechanism. Thus, for the dominant KO final 
states, $\sigma_f = b_f\, \sigma_{inc}^{exp}= b_f^{KO}\, \sigma_{inc,KO}^{exp}$, where 
$b_f^{KO}=b_f/0.953$ and $\sigma_{inc,KO}^{exp}$ is the measured inclusive KO 
cross section, as enters the definition of $R_s$.

In the present work we deduce (normalized) proton spectroscopic factors for the 
individual KO final states $f$ (with excitation energy $E^*$) from the measured
partial cross sections and the calculated single particle cross sections, as follows:
\begin{equation}
C^2 S^{exp}_{norm} = \frac{b_f^{KO} \sigma_{inc,KO}^{exp}} {R_s \sigma_f^{sp}}.
\label{eq2}
\end{equation}
In calculating these $C^2 S^{exp}_{norm}$, the $R_s$ trend from the inclusive data,
Eq.\ (\ref{eq1}), is now used for each final state and is calculated for the proton 
separation energy to that state, i.e. $S_p=S_p(g.s.)+E^*$. As we observe excited 
states with $E^*$ up to 7.5 MeV, $R_s$ varies by about 20\% over this range.
We note that if a constant (inclusive) $R_s$ value had been used
for all final states, the deduced normalized $C^2S$ values shown
in Table \ref{C2S} would still agree with those presented
within the stated errors. Our use of  $R_s$ in the denominator in Eq.\ (\ref{eq2}), means that 
we extract shell-model-like (and not suppressed) spectroscopic factors that, in 
a sum-rule limit, can be compared to the occupancies of the active orbits in a 
finite-basis shell-model calculation. Thus, as defined, the summed $C^2 S^
{exp}_{norm}$ for a given $J$ should have a maximum value of $(2J+1)$ in the 
limit that an orbital with angular momentum $J$ is fully occupied.

The $\ell$ value of the removed proton is determined by comparing measured
parallel momentum distributions ($p_{//}$) of the $^{35}$P residues to theoretical
distributions calculated with the same S-matrices and overlaps as used for the 
computation of the $\sigma_f^{sp}$~\cite{Bertulani04,Bertulani06}.  In order to 
account for broadening effects on $p_{//}$, due to the incoming beam momentum 
profile and the straggling of the secondary beams that pass through the target, 
theoretical $p_{//}$ distributions were folded with the experimental $p_{//}$ 
distribution obtained from the unreacted \nuc{36}{S} nuclei passing through the 
target. Also, the additional momentum broadening induced by the fact that the 
proton knockout reaction may happen anywhere in the target has been taken into 
account.

\section{Results and discussions}

\subsection{Proton occupancies in $^{36}$S}
As discussed earlier, the same states were populated in the knockout from
$^{36}$S to $^{35}$P and in the $^{36}$S($d,^3$He) $^{35}$P transfer~\cite{Khan85}
reactions. We now explore whether the deduced spectroscopic factors, $C^2S$,
are also comparable.

The $C^2S$ values obtained from the $(d,^3$He) proton transfer for the $2s_{1/2}$, $1d_{3/2}$, and $1d_{5/2}$ orbits, populating the 1/2$^+$, 3/2$^+$, 5/2$^+$ states in $^{35}$P \cite{Khan85} are shown in the final column of Table \ref{C2S}. The occupancy of 
the $2s_{1/2}$ orbit was about 5 times larger than that of the (in a naive picture, 
unoccupied) $1d_{3/2}$ orbit. Their summed occupancy amounts to about 2. In an 
extreme single-particle picture, this indicates that there are few excitation 
from the $2s_{1/2}$ to the $1d_{3/2}$ orbit. Five 5/2$^+$ states were proposed 
at higher excitation energy, among which the assignments of the two states at 
4474 and 7520~keV were only tentative. By adding all the (5/2$^+$) strength, 
an occupancy of 5.95 was found, very close to the sum rule of $2J+1$=6. Altogether, 
the summed occupancy of the $(2s,1d)$ orbits derived from the transfer reaction, 
$\Sigma C^2S\simeq$ 7.9 $\pm$1.6 \cite{Khan85}, exhausts the sum rule of $\Sigma 
C^2S$ = 8. As, with the DWBA model parameters used by the authors, the derived
$C^2S^{(d,^3He)}$ exhaust the truncated-basis shell-model sum rule, these
(unsuppressed) values should be compared directly with the shell-model and
with the normalized values $C^2 S^{exp}_{norm}$ from the KO analysis.
One should, however, take the $1d_{5/2}$ strength 
as maximum as the authors of Ref. \cite{Khan85} tentatively proposed to add two 
levels with $\ell$=2 to their sum (the one at 4474(21) keV with $C^2S$ =0.2, the 
other at 7520(30) keV with $C^2S$ =0.4) for which the angular distributions of 
the $^3$He could not be measured. Moreover, a 7/2$^-$ negative-parity assignment 
was later proposed in Ref.~\cite{Wide08} for the 4494~keV state. In the following 
we attempt to clarify this situation with the help of the present data set.

\begin{table*}
\caption{Energy, spin parity, partial cross-sections ($b_f^{KO} \times \sigma_{inc,KO}^{exp}$),
and normalized experimental spectroscopic factors $C^2 S^{exp}_{norm}$ of $^{35}$P final states
populated in the $^{36}$S$(-1p)$ reaction. The spectroscopic factors derived from the ($d$,$^3$He)
transfer reaction \cite{Khan85} are shown in the last column.}
\setlength{\tabcolsep}{4pt}
\begin{tabular}{ c c  c c  c  c  c c c }
\hline
E  & J$^\pi$ & $b_f^{KO}$ $^\dagger$& $b_f^{KO} \times \sigma_{inc,KO}^{exp}$
& $\sigma^{sp}$ &$R_s$ & $C^2 S^{exp}_{norm}$ $^\ddag$& $C^2S^{(d,^3He)}$ \rule[-7pt]{0pt}{20pt}\\
(keV) &&(\%)&(mb)&(mb)&\\
\hline
0 & $\frac{1}{2}^+$ & 32.9(7.5)&16.0(3.6) &13.5 &0.55(11)&2.2(7) & 1.6(3) \rule[-7pt]{0pt}{20pt} \\
2386 & $\frac{3}{2}^+$ &  8.2(3.8)& 4.0(1.8) & 10.2&0.52(10)&0.7(3)& 0.31(6) \rule[-7pt]{0pt}{20pt} \\
3860 & $\frac{5}{2}^+$ &  28.6(6.1) &13.9(3.0) &10.6 &0.49(10)&2.7(8) & 2.9(6) \rule[-7pt]{0pt}{20pt} \\
4494 & $(\frac{7}{2}^-)$ &  3.0(5)& 1.5(2)   &10.7&0.48(10)&0.30(7)  & $<$0.2  $^\star$ \rule[-7pt]{0pt}{20pt} \\
4665 & $\frac{5}{2}^+$ &  9.9(1.1)& 4.8(5) & 10.3&0.47(9)&1.0(2) & 1.1(2) \rule[-7pt]{0pt}{20pt} \\
5197 & $\frac{5}{2}^+$ &  15.2(1.5)&7.4(7) & 10.2&0.47(9)&1.5(3) & 1.4(3) \rule[-7pt]{0pt}{20pt} \\
5709 & ($\frac{1}{2}^-$) &  1.5(1.1)& 0.73(53) & 10.8&0.47(9)&0.21(16) & - \rule[-7pt]{0pt}{20pt} \\
7526 & ($\frac{1}{2}^-$) &  1.3(3)&0.63(15)& 10.2&0.44(9)&0.20(6) & $<$0.4 $^\star$ \rule[-7pt]{0pt}{20pt} \\
\hline
\end{tabular}
\\$^\dagger$ Normalized to have the sum of $(-1p)$ knockout events =100\% \\
$^\ddag$ Normalized according to Eq. (\ref{eq2}), and hence the $C^2 S^{exp}_{norm}$ sum  to $(2J+1)$
in the case of a fully occupied sub-shell.\\
$^\star$  Value obtained in \cite{Khan85} by tentatively assuming a $d_{5/2}$ proton transfer.
\label{C2S}
\end{table*}

\begin{figure} [t]
\includegraphics[width=\columnwidth] {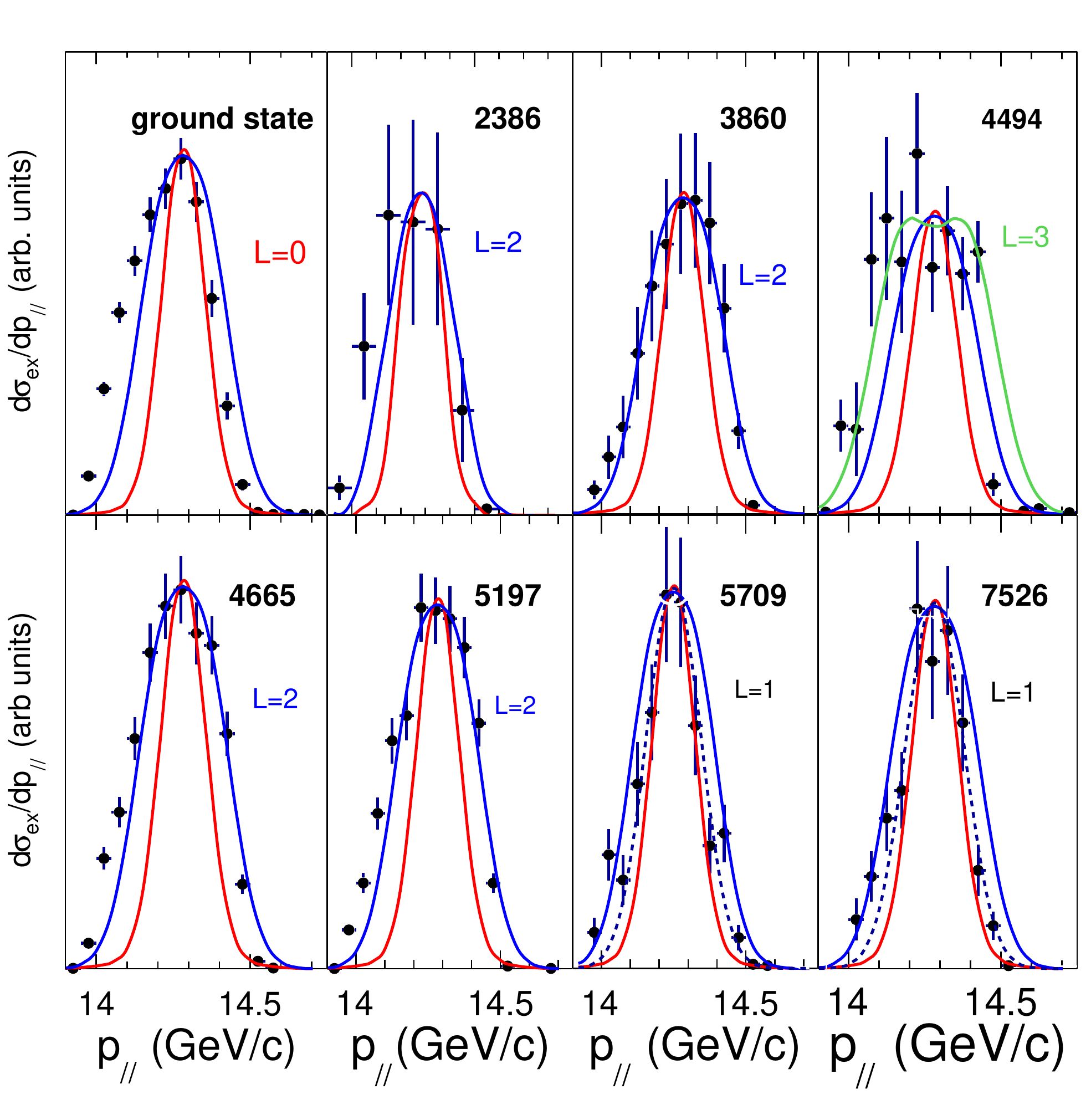}
\caption {(color on line) Experimental parallel momentum distributions for the most populated states in $^{35}$P (black crosses) are compared to calculations assuming a $\ell=0$ (red curves), $\ell=1$ (dashed blue), $\ell=2$ (blue) and $\ell=3$ (green) proton removal from the
$^{36}$S ground state.}
\label{35P-ppar}
\end{figure}

Comparisons between experimental and theoretical $p_{//}$ distributions are
shown in the panels of Fig. \ref{35P-ppar} for the states represented in the
left hand side of Fig. \ref{35P-ls} (with black lines) and populated with
$b_f$ values having a statistical significance of at least a 3 $\sigma$.
Comparison should focus on the high-momentum part of the distributions as
the low momentum part exhibits a tail, attributed to events in which energy
is dissipated in exciting the target \cite{Stroberg14}, the kinematics of
which are not correctly treated by the eikonal model. The three 5/2$^+$
excited states at 3860, 4665 and 5197 keV exhibit experimental $p_{//}$
distributions that are compatible with calculated distributions computed
for $\ell$=2 proton removal.  Though the statistics are limited, the $p_{//}$
distribution corresponding to the 4494 keV seems more consistent with an
$\ell$=3 assignment. It likely corresponds to the state observed, but
unassigned, at 4474(21) keV in Ref. \cite{Khan85}. The newly identified
state at 5709 keV and the previously unassigned state at 7520(30) keV in
\cite{Khan85} display $p_{//}$ distributions in closer agreement with
$\ell$=1, most likely corresponding to deep-hole configurations involving
the $1p_{1/2}$ orbital. No direct feeding of the state at 4101 keV is
observed, in agreement with its non-feeding in the $(d,^{3}$He) reaction
\cite{Khan85}. Finally, the $p_{//}$ distribution leading to the ground
state is in agreement with an $\ell$=0 proton removal and its corresponding
$J^\pi$ = 1/2$^+$ assignment. As the population fraction $b_f$ for the
ground state is obtained by subtracting the contributions of all excited
states, its uncertainty is large. The $\ell$ assignments for the most
strongly populated states agree with those from the transfer reaction
\cite{Khan85}. New tentative assignments are proposed in Table~\ref{C2S}
for the previously unassigned states at 4494 and 7520 keV and the newly
identified state at 5709 keV.

The measured inclusive cross section that leads to $^{35}$P, from incoming
$^{36}$S projectiles, is 51(1) mb, of which 95.3 \% corresponds to the direct
proton knockout (KO) process. The corresponding partial cross sections, $b_f^{KO}
\times \sigma_{inc,KO}^{exp}$, as well as the derived $C^2 S^{exp}_{norm}$ values,
are reported in Table \ref{C2S}. All of these $C^2 S^{exp}_{norm}$ values are,
within the error bars, consistent with those for $C^2S^{(d,^3He)}$ \cite{Khan85}. 
Thus, despite the use of the different reaction mechanisms and theoretical models, 
consistent proton spectroscopic factors could be derived using the one-proton 
knockout and the ($d,^3$He) reaction probes.

\begin{figure} [t]
\includegraphics[height=13cm, width=6cm] {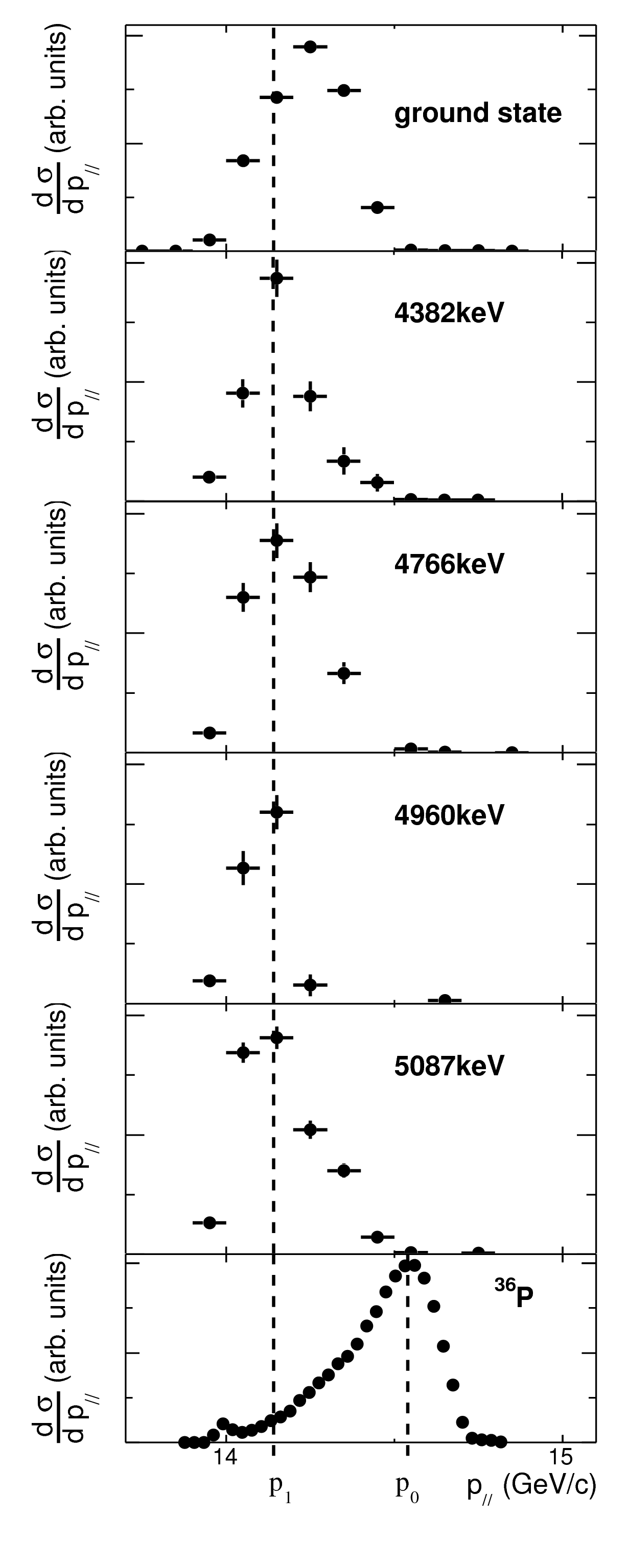}
\caption {Measured parallel momentum distributions for the ground state and the
 states at 4382, 4766, 4960 and 5096 keV (displayed in red in the level scheme)
 of $^{35}$P. The bottom panel shows that for $^{36}$P nucleus, as produced from
 \nuc{36}{S}. The vertical dashed line, crossing all panels, represents the
 centroid of the momentum distribution of $^{36}$P nucleus, $p_0$, shifted by
 the evaporation of a neutron, i.e. $p_1$=(A-1)/A $p_0$. It is seen that the
 distributions for all states labeled red in the level scheme are shifted
 relative to that of the $^{35}$P ground state, that is produced via direct
 one-proton knockout.}
\label{35P-ppar-shift}
\end{figure}

\subsection{States in $^{35}$P produced by another mechanism} \label{TS}

As shown in Fig. \ref{35P-ppar-shift}, the $p_{//}$ centroids for those $^{35}$P
final states indicated in red on the right hand side of the level scheme
in Fig. \ref{35P-ls}, are systematically shifted to a lower momentum compared
to all distributions earlier attributed to the one-proton knockout reaction.
It is interesting to compare these centroids with that obtained for $^{36}$P
produced from \nuc{36}{S} in a \nuc{9}{Be}-induced nucleon-exchange reaction
~\cite{Gade09}. In Fig. \ref{35P-ppar-shift}, it is shown that the momentum-shifted
centroids of $^{35}$P match the one deduced from the experimental $p_{//}$
distribution of $^{36}$P after being scaled by a factor of 35/36, that would
correspond to the evaporation of one neutron following nucleon exchange into
\nuc{36}{P}. Having similar cross sections, the states in $^{36}$P ($\sigma$=
1.5(5) mb) and the momentum-shifted ones in $^{35}$P ($\sigma$=2.40(4) mb),
are likely populated by the same nucleon-exchange mechanism. It is conceivable
that these states are reached either by: (i) a two-step (TS) process involving
a proton knockout and a neutron pickup, or (ii) a charge-exchange (CE) reaction.
Either way, bound or unbound states may be produced in $^{36}$P, leading to
characteristic $\gamma$ rays in $^{36}$P or decay to $^{35}$P, respectively.
These states could also be produced, as proposed in Ref.~\cite{Stroberg14},
by a (iii) TS reaction pathway in which the $^{36}$S core is excited by inelastic
scattering with proton removal from the excited state. These three possibilities
are examined in the following.

When assuming that the knockout and pickup reactions have a similar dependence
on impact parameter, the cross section for a TS process can be derived from
the product of the two individual interaction probabilities. A knockout cross
section of about 50 mb was derived from our experiment, while a neutron pickup
cross section of about 2.5 mb was measured at similar beam energy for the
$sd$-shell nucleus $^{22}$Mg~\cite{Gade11}. It follows that the TS mechanism
would have a rather low cross section, of the order of 0.10 mb, that is a
factor of 20 lower than we observe for the sum over all states with shifted
momentum distributions. It is therefore likely that a more direct process is
needed to account for the measured cross section.

\begin{figure} [t]
\includegraphics[width=\columnwidth] {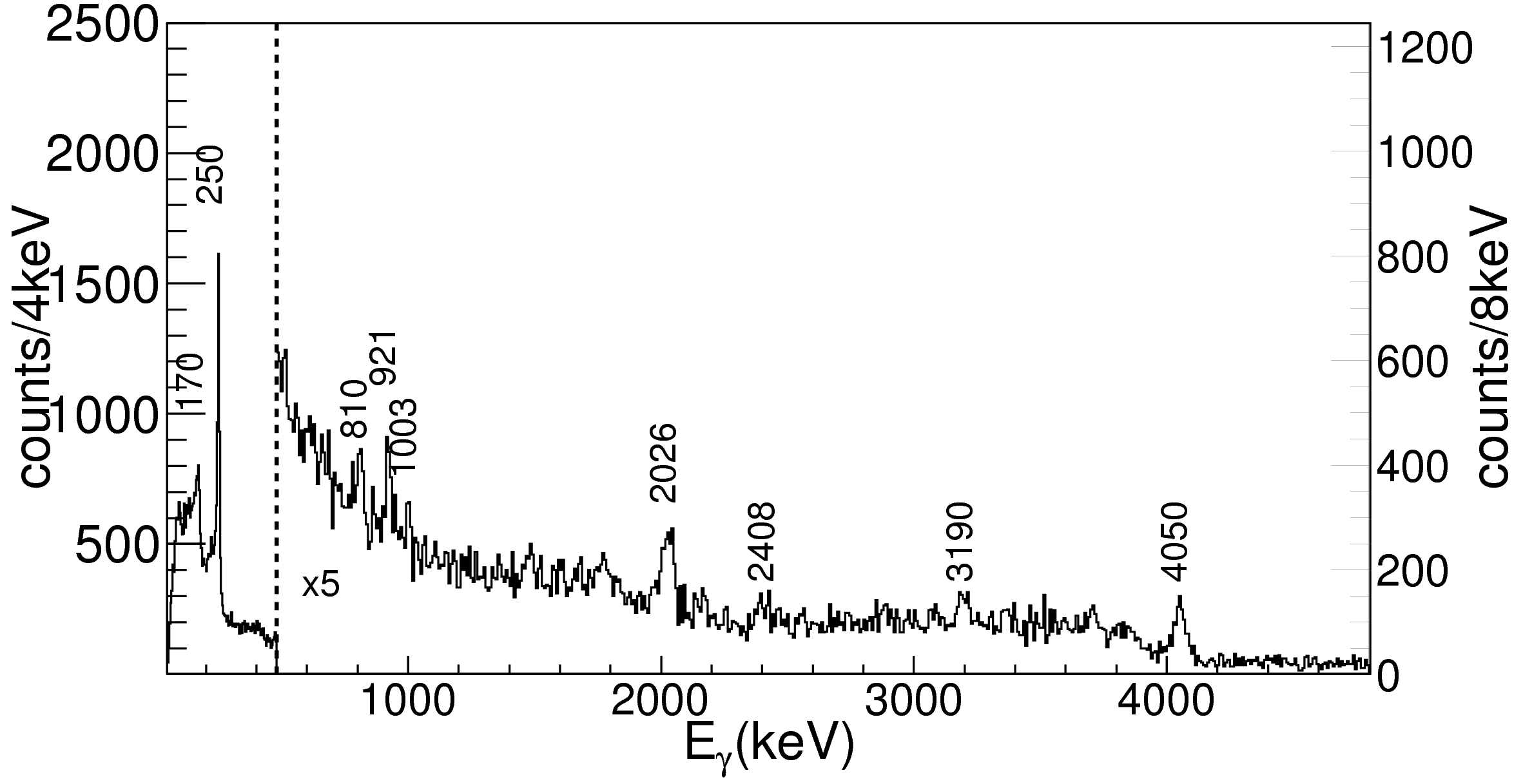}
\caption {Doppler-corrected $\gamma$-decay spectrum for the $^{36}$P nucleus
  populated from $^{36}$S in a nucleon-exchange reaction. }
\label{36P}
\end{figure}

Charge-exchange reactions occur more favorably with an allowed GT operator
($\Delta L$=0,  $\Delta J$=$0,\pm $1, $\Delta \pi$=no)  that would lead to the
production of 1$^+$ states in $^{36}$P starting from the 0$^+$ ground state of
$^{36}$S. Higher momentum transfers are also observed when very strict
kinematic condition for the reaction as well as suitable target excitations
cannot be selected, as in the present experiment. Moreover, having  a closed
neutron $sd$ shell, the exchange of an $sd$ proton to an $sd$
neutron orbital that would lead to 1$^+$ states in $^{36}$S is strongly hindered
by Pauli blocking. Rather, transfer to the $fp$ or to the highest $g-sd$ orbits
must become largely competitive. The non-feeding of any of the known 1$^+$
states in $^{36}$P but the population of the negative parity states 4$^-$ (g.s.),
3$^-$ (250~keV) and of the tentatively assigned 2$^-$ (425 keV) through
low-energy charge-exchange reactions of Ref. \cite{Fif93} are in line with the
above assumptions. In the present experiment, several $\gamma$ rays are observed
in coincidence with $^{36}$P (see Fig. \ref{36P}). However, none of them arise
from the deexcitation of the 1$^+$ states at 1303~keV and 2281~keV that were
populated in the $\beta$ decay of $^{36}$Si to $^{36}$P \cite{Dufour2} and whose
decay occurs through $\gamma$ rays of 878~keV and (934, 977, 1858~keV), respectively,
as in Ref.~\cite{Fif93}. Even if the present statistics does not allow extracting
$\gamma$-$\gamma$ coincidences, the observation of transitions at 250~keV and
about 170~keV suggest the feeding of the 3$^-$ and 2$^-$ states, respectively.
Other transitions are observed, but none (except perhaps the 2020(20)~keV from
\cite{Orr88}) were reported in previous works. Interestingly, one transition is
seen at 4050~keV, i.e. about 600~keV above the neutron emission threshold of
3.465~MeV. This state seems to decay preferentially through a $\gamma$-ray
transition rather than neutron decay to the 1/2$^+$ ground state of
$^{35}$P.

The last hypothesis is that these momentum-shifted distributions result from the
proton knockout from a state excited in inelastic scattering. Such a TS reaction
was discussed as one possibility to account for the observation of a
$\gamma$-decaying neutron-unbound state in $^{35}$Si at 3.611~MeV
($S_n$=2.470(40)~MeV) populated in the $^{36}$Si
$(-1n)$ reaction~\cite{Stroberg14}. As shown in Fig. 4 of Ref.~\cite{Stroberg14}, the
momentum distribution associated with the production of this unbound state
appears to lie dominantly in the low-momentum tail of the $^{35}$Si
distribution, possibly indicating that its centroid is shifted to lower
momentum. The production of such a state was
evaluated in Ref. \cite{Stroberg14} to be of 0.8(2) mb, a value that is
comparable to the partial cross sections for the states which are observed with
a shifted momentum distribution in our work. A total cross section of 2.4(4)
mb is indeed obtained by summing over the four states represented in red in
Fig. \ref{35P-ls}.  Contrary to the case presented in Ref. \cite{Stroberg14},
the states here possibly produced by the same TS mechanism are below the neutron
separation energy. This could
be due to the fact that the neutron emission threshold of $^{35}$P
($S_n$=8.380 MeV) is much higher than the one of  $^{35}$Si ($S_n$=2.470 MeV).

\section{Conclusions}

The spectroscopy of $^{35}$P was investigated with an intermediate-energy
one-proton knockout reaction using  a secondary beam of $^{36}$S interacting
with a $^{9}$Be foil at 88 MeV/u mid-target energy. The $^{35}$P nuclei were
selected by the S800 spectrograph at NSCL and identified in the spectrograph's 
focal plane in coincidence with prompt $\gamma$ rays detected in the GRETINA 
segmented Ge array around the target. The level scheme of $^{35}$P has been 
established up to about 7.5 MeV from $\gamma$-$\gamma$ coincidences and relative 
$\gamma$ intensities. Spins and parities for most of the populated states were 
proposed from a comparison between calculated and measured $\gamma$-gated 
momentum distributions for the population of individual final states in $^{35}$P. 
Spectroscopic factor values, $C^2S$, were derived from partial cross sections 
for the 1/2$^+$ ground state and the excited states (3/2$^+$, 5/2$^+$ and 
tentatively 1/2$^-$) in $^{35}$P. An inclusive cross section of 51(1) mb was 
measured for \nuc{35}{P} produced from \nuc{36}{S}. The extracted summed $\sum
C^2S$ values agree with expected shell-model occupancies of the protons in 
the $2s_{1/2}$ ($\sum C^2S$=2.2(7)), $1d_{3/2}$ (0.7(3)), $1d_{5/2}$ (5.2(9)) 
orbitals, while only a fraction (0.41(17)) of the strength corresponding to the deeply-bound 
$1p_{1/2}$ orbit is tentatively observed for two states above 5.6 MeV.

The present results were compared to those obtained for $^{35}$P using the
$^{36}$S($d,^3$He)$^{35}$P transfer reaction at low energy. Remarkable agreement
is found for the proposed level scheme populated, the spin assignments (except
for one state) and the deduced $C^2S$ values. Since obtained from high-resolution
$\gamma$-ray spectroscopy, the excitation energies of the presently identified
states in $^{35}$P are more accurate than those obtained in the ($d,^3$He)
reaction \cite{Khan85}. They agree with the values obtained in a multi-nucleon 
transfer reaction \cite{Wide08}, for which several states were observed in common. 
The present sensitivity of the knockout reaction, with a projectile beam of 2
$\cdot$10$^5$ pps, matches the one obtained in the $(d,^3$He) reaction
with a $d$ beam on a stable $^{36}$S target. This reinforces the enormous
potential of this experimental technique in extracting level schemes, orbital
angular momenta, and $C^2S$ values.

Besides the observation of states that were expected to be produced in the
knockout of protons from the $p-sd$ shell, other states with likely high spin
value ($J \geqslant$ 5/2) and negative parity are observed as well, with a
summed partial cross section of about 2.4 mb. Owing to the facts that these
states were not observed in the ($d,^{3}$He) reaction and that they exhibit
parallel-momentum distributions that are down-shifted as compared to the ones
from knockout, we propose that they are produced by another mechanism that could
be either a charge-exchange or a two-steps mechanism in which a core excitation
is followed by a proton knockout. While so far we cannot clearly identify
which of these two mechanisms is the most probable, it interestingly leads to
the production of states most likely corresponding to neutron-core
excitations. This feature may be very interesting to single-out intruder states
belonging to the so-called island of inversion, as evoked in
Refs. \cite{Stroberg14,Zegers10}.

\acknowledgments {\small}
This work is supported by the National Science Foundation (NSF) 
under Grant Nos. PHY-1102511 and PHY-1306297 and by the Institut Universitaire de France. GRETINA was funded by the US DOE - Office of Science. Operation of the array at NSCL is supported by NSF under  Cooperative Agreement PHY-1102511 (NSCL) and DOE under grant DE-AC02-05CH11231  (LBNL). O.S wish to thank T. Duguet for fruitful discussions. J.A.T acknowledges support of the Science and Technology Facility Council (UK) grant ST/L005743.

\end{document}